\documentclass{article}
\def\today{\number\day\space\ifcase\month\or
  January\or February\or March\or April\or May\or June\or
  July\or August\or September\or October\or November\or December\fi
  \space\number\year}
\catcode`\@=11
\renewcommand{\ps@plain}{%
  \renewcommand{\@oddhead}{}%
  \renewcommand{\@evenhead}{}%
  \renewcommand{\@evenfoot}{\hfil\textrm{\thepage}\hfil\today}%
  \renewcommand{\@oddfoot}{\@evenfoot}}
\catcode`\@=12

\usepackage{amssymb}
\usepackage{proof}
\long\def\omitthis#1{}
\newtheorem{thm}{Theorem}[section]

\newcommand{\qed}{\hbox{\vrule width1.0ex height1.0ex}}
\newenvironment{prf}{{\bf Proof:}\hspace*{0.5em}}{\hspace*{\fill}\qed}

\newcommand{\implies}{\Rightarrow}
\newcommand{\litset}[1]{\{#1\}}
\newcommand{\dom}[1]{{\it dom}(#1)}
\newcommand{\fv}[1]{{\it fv}(#1)}
\newcommand{\bnfdef}{\mathrel{::=}}
\newcommand{\bnfalt}{\mathrel|}
\newcommand{\subst}[2]{#1\{#2\}}
\newcommand{\MVTL}{\lambda_\aleph}
\newcommand{\kw}[1]{{\sf#1}}
\newcommand{\fx}{\chi}
\newcommand{\fxT}{\kw{T}}
\newcommand{\fxP}{\kw{P}}
\newcommand{\fxN}{\kw{N}}
\newcommand{\fxR}{\kw{R}}
\newcommand{\fxW}{\kw{W}}
\newcommand{\fxIO}{\kw{IO}}
\newcommand{\fxRV}{\kw{RV}}
\newcommand{\fxA}{\kw{A}}
\newcommand{\fxle}{\sqsubseteq}
\newcommand{\fxmeet}{\sqcap}
\newcommand{\fxseq}{\mathbin;}
\newcommand{\dtrue}{\kw{T}}
\newcommand{\dfalse}{\kw{F}}
\newcommand{\ddecidable}{\kw{D}}
\newcommand{\falses}{\kw{falses}}
\newcommand{\anys}{\kw{anys}}
\newcommand{\ints}{\kw{ints}}
\newcommand{\tabs}{\kw{tabs}}
\newcommand{\funs}{\kw{funs}}
\newcommand{\ptrs}{\kw{ptrs}}
\newcommand{\ptrto}[1]{\kw{ptr}(#1)}
\newcommand{\uop}[1]{{\it uop}\;#1}
\newcommand{\bop}[2]{#1\mathbin{{\it bop}}#2}
\newcommand{\cop}[2]{#1\mathrel{{\it cop}}#2}
\newcommand{\TABLE}[1]{\langle#1\rangle}
\newcommand{\tabe}[3]{#2\mapsto#1=#3}
\newcommand{\tabeV}[2]{#1\mapsto#2}
\newcommand{\arrL}[3]{\kw{arr}[#1]#2\mapsto#3}
\newcommand{\dk}{{\it dk}}
\newcommand{\dkC}{-}
\newcommand{\dkI}{\circ}
\newcommand{\dkR}{\mathord\ge}
\newcommand{\dkS}{\mathord\le}
\newcommand{\dkof}[1]{{\it dk}(#1)}
\newcommand{\funL}[6]{\kw{fn}^{#3}(#1=#2[#4])[#5]#6}
\newcommand{\funA}[8]{\kw{fn}^{\dkC}\{#1=#2;#3\}(#4=#5[#6])[#7]#8}
\newcommand{\len}[1]{\kw{len}(#1)}
\newcommand{\appE}[2]{#1(#2)}
\newcommand{\appF}[2]{#1[#2]}
\newcommand{\from}[1]{\mathord:#1}
\newcommand{\ptrnew}[2]{\kw{new}(#1,#2)}
\newcommand{\ptrrd}[1]{\mathord!#1}
\newcommand{\ptrwr}[2]{#1\mathbin{:=}#2}
\newcommand{\intin}{\kw{in}}
\newcommand{\intout}[1]{\kw{out}(#1)}
\newcommand{\unify}[2]{#1\mathbin{==}#2}
\newcommand{\join}[2]{#1\mathbin|#2}
\newcommand{\LET}[3]{\kw{let}(#1=#2)#3}
\newcommand{\LETREC}[2]{\kw{letrec}(#1)#2}
\newcommand{\IF}[4]{\kw{if}\;(#1=#2)\;#3\;\kw{else}\;#4}
\newcommand{\IFM}[5]{\kw{if}\;(#1=#2)\;#3\;\kw{else}_{#4}\;#5}
\newcommand{\stage}[4]{\kw{S}_{#1,#2}(#3,#4)}
\newcommand{\fxthen}[2]{#1\vartriangleright#2}
\newcommand{\hl}{{\it hl}}
\newcommand{\pl}{{\it pl}}
\newcommand{\ptrc}[3]{(#1;#2;#3)}
\newcommand{\env}{\sigma}
\newcommand{\FRAME}[3]{#1(#2)^{#3}}
\newcommand{\AEQ}{{\it ae}}
\newcommand{\chk}[5]{#1\mathrel{\in_{#2}}#3\vartriangleright#4\;\kw{else}\;#5}
\newcommand{\clos}[2]{(#1;#2)}
\newcommand{\HH}{{\it HH}}
\newcommand{\PH}{{\it PH}}
\newcommand{\ms}[5]{(#1;#2;#3;#4;#5)}
\newcommand{\Ain}[1]{\kw{I}(#1)}
\newcommand{\Aout}[1]{\kw{O}(#1)}
\newcommand{\AT}{\kw{T}}
\newcommand{\Anew}{\kw{N}}
\newcommand{\Ard}{\kw{R}}
\newcommand{\Awr}{\kw{W}}
\newcommand{\stepsto}[1]{\stackrel{#1}{\mapsto}}
\newcommand{\stepstoF}{\mapsto\kw{F}}
\newcommand{\stepstoE}{\mapsto\kw{E}}
\newcommand{\evalsto}{\Downarrow}
\newcommand{\diverges}{\Uparrow}
\newcommand{\evalstoE}{\Downarrow\kw{E}}
\newcommand{\rn}[1]{\hbox{\sc#1}}
\newcommand{\emptyseq}{\epsilon}
\newcommand{\hole}{\mathord{[]}}
\newcommand{\substhole}[2]{#1[#2]}
\newcommand{\Hisint}[1]{\mathord{{\it int}?}(#1)}
\newcommand{\Hisnat}[1]{\mathord{{\it nat}?}(#1)}
\newcommand{\Histab}[1]{\mathord{{\it tab}?}(#1)}
\newcommand{\Histabf}[2]{\mathord{{\it tab}?_{\litset{#1}}}(#2)}
\newcommand{\Hisarr}[1]{\mathord{{\it arr}?}(#1)}
\newcommand{\Hisfun}[1]{\mathord{{\it fun}?}(#1)}
\newcommand{\Histyp}[1]{\mathord{{\it typ}?}(#1)}
\newcommand{\Hisapp}[1]{\mathord{{\it app}?}(#1)}
\newcommand{\Hisptr}[1]{\mathord{{\it ptr}?}(#1)}
\newcommand{\evalV}[5]{#1;#2;#3\Downarrow#4;#5}
\newcommand{\evalVE}[2]{#1;#2\Downarrow\kw{E}}

\title{Formalisation of the $\MVTL$ Runtime}
\author{
\begin{tabular}{c}
Neal Glew \\
Intel Labs \\
{\small\tt aglew@acm.org}
\end{tabular}
\and
\begin{tabular}{c}
Tim Sweeney \\
Epic Games \\
{\small\tt tim.sweeney@epicgames.com}
\end{tabular}
\and
\begin{tabular}{c}
Leaf Petersen \\
Intel Labs \\
{\small\tt leaf.petersen@intel.com}
\end{tabular}
}
\begin{document}
\maketitle
\section{Introduction}

Strong static type systems eliminate errors.  Compilers refuse to
compile programs that fail the type discipline, guaranteeing that
those programs that pass the discipline are free of a certain category
of errors.  For mainstream programming languages that category of
errors includes applying primitive operations to values of
inappropriate type, calling something that is not a function,
subscripting something that is not an array, and so on.  However, it
does not include errors such as subscripting outside the bounds of an
array or creating red-black trees that do not satisfy the red-black
invariant---these errors can only be checked dynamically.  Eliminating
these errors statically is desirable.

Dependent type systems can be used to eliminate such errors.  In
previous work~\cite{mvtl-overview} we introduced a novel approach to
dependent typing, a language we call $\MVTL$.  Our language has two
key features that combine to make it suitable for dependent typing.
First, it does not have separate languages for expressions, types,
kinds, and sorts, it just has a single language of terms.  Second,
terms specify sets of values rather than single values, as in most
languages.  Terms that specify singleton sets are used to specify
computations, and terms that specify sets of arbitrary cardinality are
used to express type information.  Since the term language is as rich
as most expression languages, quite interesting types can be
expressed.  In particular, variables can be used to refer to other
arguments or other components of a data structure, thus allowing for
dependent typing.  For more details on how this works, see our
previous paper.

Our previous paper describes the motivation and intuitions behind
$\MVTL$, illustrates how it can used to do dependent typing, gives
examples of the constructs in the language, and then informally
describes what the language is precisely.  The paper sketches a
\emph{set-theoretic semantics}, a form of denotational semantics, for
$\MVTL$.  This semantics defines for each term what set of values it
specifies.  However, the semantics is not directly realisable on a
machine.  Instead, a subset of the full term language is selected for
which an implementation is realisable.  That implementation is called
the \emph{runtime}, and the paper shows part of a formalisation of the
runtime as an operational semantics for $\MVTL$.  To ensure that
programs are in the subset the runtime is intended for, a static
checking process called the \emph{verifier} checks for and rejects
programs that not.  Since we intend for $\MVTL$ to be strongly
statically typed, the verifier also checks for an rejects programs
that might commit type errors, such as those mentioned above.

This technical report completely formalises the runtime, filling in
all the details not covered in our previous paper.  It does not
motivate nor describe the design behind the runtime, rather it just
states the formalisation.  For details on why the runtime is
structured the way it is, why it does correctly implements the
set-theoretic semantics on the subset it is design for, or what
precisely that subset is and how the verifier checks for it, see our
previous paper.

The actual language formalised is a comprehensive core
language, and is intended to cover most of the constructs in a
realistic language after it has been desugared and elaborated.  Such a
language would probably have a few more primitive types, such as
single- and double-precision floating point (it might also have rational
numbers rather than integers), a notion of names, and a more general
version of tables including, probably, a few more operations to
combine or update tables.  Otherwise we believe it covers everything
of interest.

The formalisation takes the form of an abstract machine and reduction
relations for it.  We make errors explicit---the abstract machine
steps to an error if the next operation to perform is type
erroroneous.  The verifier should ensure the runtime does not step to
an error and is formalised elsewhere (currently we are working on this
and anticipate publishing a technical report at some stage).

\section{Syntax}

In this section we describe the term language itself.  First we
describe a couple of preliminaries: effects and decidabilities; next
we describe the term language; finally we say what a program is.

\subsection{Effects}

We use an effects lattice to describe the possible computational
effects that a term might commit when evaluated.  The lattice is the
product of several two point lattices that correspond to the primitive
effects of partiality, pointer creation, pointer dereference, pointer
update, and IO.  Specifically, an effect, ranged over by metavariable
$\fx$, is a subset of the set $\litset{\fxP,\fxN,\fxR,\fxW,\fxIO}$.
Ordering in the lattice is just subset and is denoted $\fxle$.
Join and meet in the lattice are just union and intersection.
Sequencing of two effects $\fx_1\fxseq\fx_2$ is simply join.  The least
effect is the emptyset and is also denoted $\fxT$; the greatest effect
is the whole set and is also denoted $\fxA$.  The reversible effects
are denoted $\fxRV$, currently $\fxRV=\litset{\fxP,\fxN,\fxR,\fxW}$.

\subsection{Decidabilities}

A term specifies a set of \emph{outcomes}, namely, values, divergence,
or type errors.  A term is \emph{inhabited} if it has at least one
outcome (specifies a set of at least one outcome), and is
\emph{uninhabited} if it has no outcomes (specifies the empty set of
outcomes).  Decidabilities indicate whether a term is definitely
inhabited, definitely uninhabited, or of unknown inhabitance.
Specifically, a decidability is ranged over by metavariable $d$ and is
given by this grammar:
\[
\begin{array}{lcl}
d &\bnfdef& \dtrue \bnfalt \dfalse \bnfalt \ddecidable \\
\end{array}
\]

\subsection{Terms}

The terms, ranged over by metavariable $t$, are given by this grammar:
\[
\begin{array}{lcl}
t &\bnfdef&
  x \bnfalt \falses \bnfalt \anys \bnfalt {}\\&&
  i \bnfalt \ints \bnfalt \uop{t} \bnfalt \bop{t_1}{t_2} \bnfalt \cop{t_1}{t_2} \bnfalt {}\\&&
  \TABLE{\tabe{x_1}{i_1}{t_1},\ldots,\tabe{x_n}{i_n}{t_n}} \bnfalt \arrL{t_1}{x}{t_2} \bnfalt \tabs \bnfalt {}\\&&
  f \bnfalt \funs \bnfalt {}\\&&
  \len{t} \bnfalt \appE{t_1}{t_2} \bnfalt \appF{t_1}{t_2} \bnfalt \from{t} \bnfalt {}\\&&
  \ptrnew{t_1}{t_2} \bnfalt \ptrrd{t} \bnfalt \ptrwr{t_1}{t_2} \bnfalt \ptrto{t} \bnfalt \ptrs \bnfalt {}\\&&
  \intin \bnfalt \intout{t} \bnfalt {}\\&&
  \unify{t_1}{t_2} \bnfalt (\join{t_1}{t_2}) \bnfalt {}\\&&
  \LET{x}{t_1}{t_2} \bnfalt \LETREC{x_1=v_1,\ldots,x_n=v_n}{t} \bnfalt
  \IF{x}{t_1}{t_2}{t_3} \bnfalt {}\\&& 
  \stage{\fx}{d}{t_1}{t_2} \bnfalt \fxthen{\fx}{t} \\
{\it cop} &\bnfdef&
  \mathord{<} \bnfalt \mathord{\le} \bnfalt \mathord{>} \bnfalt \mathord{\ge} \bnfalt \mathord{\ne} \\
\dk &\bnfdef& \dkC \bnfalt \dkI \bnfalt \dkR \bnfalt \dkS \\
f &\bnfdef& \funL{x}{t_1}{\dk}{\fx_1}{\fx_2}{t_2} \bnfalt \funA{x_1}{t_1}{t_2}{x_2}{t_3}{\fx_1}{\fx_2}{t_4} \\
v &\bnfdef&
  \TABLE{\tabeV{i_1}{x_1},\ldots,\tabeV{i_n}{x_n}} \bnfalt f \bnfalt \ptrnew{t}{x} \\
\end{array}
\]
where metavariable $x$ ranges over variables, $i$ over integers, $n$
over natural numbers, ${\it uop}$ over some collection of unary
operations on the integers, and ${\it bop}$ over some collection of
binary operations on the integers.

A quick explanation of what the terms means.  We have variables, the
uninhabited term $\falses$, the term for all values $\anys$, integers,
the term for all integers $\ints$, unary operations on integers
$\uop{t}$, binary operations on the integers $\bop{t_1}{t_2}$,
comparison of integers $\cop{t_1}{t_2}$, fixed table construction
$\TABLE{\tabe{x_1}{i_1}{t_1},\ldots,\tabe{x_n}{i_n}{t_n}}$, array
creation $\arrL{t_1}{x}{t_2}$, the term for all tables $\tabs$,
functions, the term for all functions $\funs$, length of an array
$\len{t}$, error application $\appE{t_1}{t_2}$, failing application
$\appF{t_1}{t_2}$, from $\from{t}$, pointer creation
$\ptrnew{t_1}{t_2}$, pointer dereference $\ptrrd{t}$, pointer update
$\ptrwr{t_1}{t_2}$, the term for all pointers of a particular type
$\ptrto{t}$, the term for all pointers $\ptrs$, input of an integer
$\intin$, output of an integer $\intout{t}$, unify $\unify{t_1}{t_2}$,
join $\join{t_1}{t_2}$, lets, recursive lets, conditionals, stage
$\stage{\fx}{d}{t_1}{t_2}$, and effects then $\fxthen{\fx}{t}$, a term
that commits arbitrary specified effects before acting like another
term.

In a little more detail: The term $\arrL{t_1}{x}{t_2}$ evaluates $t_1$
to a natural number, an array of that length is created, and then $x$
is bound consequtively to the indices of the array starting at zero
and $t_2$ is evaluated to initialise that element of the array.
Ordinary functions have the form
$\funL{x}{t_1}{\dk}{\fx_1}{\fx_2}{t_2}$; here $\dk$ is the domain
kind, $\dkC$ means contravariant, $\dkI$ means invariant, $\dkR$ means
the domain is at least $t_1$ and the function acts like $t_2$ over the
whole of its domain (not just $t_1$ as in contravariant functions),
$\dkS$ means the domain is at most $t_1$ and the function acts like
$t_2$ over its actual domain; $\fx_1$ is the effects of the domain and
$\fx_2$ is the effects of the range, they may be greater than the
actual effects committed during runtime.  All quantified contravaraint
functions have the form
$\funA{x_1}{t_1}{t_2}{x_2}{t_3}{\fx_1}{\fx_2}{t_4}$; here $x_1$ is all
quantified with domain $t_1$.  Term $t_2$ is the runtime instantiation
term, with $x_2$ bound to the actual argument $t_2$ should correctly
compute any parts of $x_1$ that are used in $t_4$ during runtime
execution.  Error application, $\appE{t_1}{t_2}$, is an error if $t_2$
is not in the domain of $t_1$; the runtime does not check this (as
with most operational semantics), but assumes the argument is in
domain and proceeds, the verifier should ensure this is correct.
Failing application, $\appF{t_1}{t_2}$, fails if $t_2$ is not in the
domain of $t_1$; the runtime does check this (for invariant functions)
and fails appropriately.  Pointer creation, $\ptrnew{t_1}{t_2}$, takes
a multivalued term $t_1$ for the type of the contents of the pointer,
the runtime ignores this, but the verifier does not, and the initial
value $t_2$ of the contents of the pointer, and creates a new pointer.
Pointer types, $\ptrto{t}$, have a multivalued term $t$ for the
possible values that the pointer points to.  The conditional
$\IF{x}{t_1}{t_2}{t_3}$ tests the inhabitance of $t_1$, if it is
inhabited it binds the value to $x$ and executes $t_2$, otherwise it
executes $t_3$.  The stage $\stage{\fx}{d}{t_1}{t_2}$ just executes
$t_2$, the rest is for the verifier.  The effects then term,
$\fxthen{\fx}{t}$, is for the verifier and it is an error to ever try
to run it.

We impose a few constraints on syntax.  In
$\TABLE{\tabe{x_1}{i_1}{t_1},\ldots,\tabe{x_n}{i_n}{t_n}}$, the integers $i_1$, \ldots,
$i_n$ must be distinct.

Variables alpha-vary, and we consider terms equal up to
alpha-equivalence.  Scoping is as follows: in
$\TABLE{\tabe{x_1}{i_1}{t_1},\ldots,\tabe{x_n}{i_n}{t_n}}$ each $x_i$ binds in
$t_{i+1}$, \ldots, $t_n$; in $\LETREC{x_1=v_1,\ldots,x_n=v_n}{t}$ each
$x_i$ binds in $v_1$, \ldots, $v_n$ and $t$; in
$\funA{x_1}{t_1}{t_2}{x_2}{t_3}{\fx_1}{\fx_2}{t_4}$, $x_1$ binds in
$t_3$ and $t_4$ and $x_2$ binds in $t_2$ and $t_4$; the $x$ in
$\arrL{t_1}{x}{t_2}$ binds in $t_2$, in $\LET{x}{t_1}{t_2}$ binds in
$t_2$, in $\IF{x}{t_1}{t_2}{t_3}$ binds in $t_2$, and in
$\funL{x}{t_1}{\dk}{\fx_1}{\fx_2}{t_2}$ binds in $t_2$.

\subsection{Programs}

A program is simply a closed term that computes to an empty table.

\section{Abstract Machine}

The abstract machine uses indirection to represent values, making
cyclic structures and sharing explicit; specifically a head heap maps
head labels to heads, which represent all the values in the runtime.
A pointer heap is used to store the current values of each pointer,
these map pointers to their current contents, which are triples of an
environment, a term, and a head label.  The environment and term
provide a type annotation, the type the pointer points to, and this
annotation is ignored by the runtime, but is useful for proving type
safety; it can be erased in an actual implementation.  The head label
is the current value pointed to.

The abstract machine uses an explicit closure representation rather
than a substitution semantics.  It also tracks allowed effects so that
effect-annotation violations can be flagged as errors.

As explained in our summary paper, the runtime runs terms in two
modes, generate and test.  The runtime uses a particular construct to
represent test mode and everything else is generate mode.

The abstract machine uses an extended term syntax, it includes a form
of conditionals that records the pointer heap when it begins for
restoration if the condition fails, head labels, frames (execute a
given term in a different environment and also constraint the allowed
effects), and a test form that implements testing mode.

The syntax definitions are:
\[
\begin{array}{lcl}
t &\bnfdef&
  \cdots \bnfalt
  \IFM{x}{t_1}{t_2}{\PH}{t_3} \bnfalt \hl \bnfalt \FRAME{\env}{t}{\fx} \bnfalt \chk{\hl}{\AEQ}{t_1}{t_2}{t_3} \\
\env &\bnfdef& x_1=\hl_1,\ldots,x_n=\hl_n \\
h &\bnfdef&
  i \bnfalt \TABLE{\tabeV{i_1}{\hl_1},\ldots,\tabeV{i_n}{\hl_n}} \bnfalt \clos{\env}{f} \bnfalt \pl \\
\HH &\bnfdef& \hl_1=h_1,\ldots,\hl_n=h_n \\
\PH &\bnfdef& \pl_1=\ptrc{\env_1}{t_1}{\hl_1},\ldots,\pl_n=\ptrc{\env_n}{t_n}{\hl_n} \\
M &\bnfdef& \ms{\HH}{\PH}{\env}{\fx}{t} \\
\end{array}
\]
where $\AEQ$ ranges over sets of pairs of head labels.

To explain the additional terms a little more: Term
$\IFM{x}{t_1}{t_2}{\PH}{t_3}$ tests the inhabitance of $t_1$ and if it
is inhabited, binds the value to $x$ and executes $t_2$, otherwise it
restores the pointer heap to $\PH$ and executes $t_3$.  Term
$\FRAME{\env}{t}{\fx}$ executes $t$ but in environment $\env$ rather
than the one from the machine state and with allowed effects $\fx$.
Term $\chk{\hl}{\AEQ}{t_1}{t_2}{t_3}$ tests if value $\hl$ is one of
the possible values of $t_1$ and if so executes $t_2$ otherwise it
executes $t_3$, where $\AEQ$ records assumed equalities amongst head
labels used to detect cycles.

We impose a few constraints on syntax.  Specifically, in
$\chk{\hl}{\AEQ}{t_1}{t_2}{t_3}$, $t_1$ must be either a head label or
in the source syntax not the extended syntax, and both $t_2$ and $t_3$
must be closed (perhaps by using a frame to bind the otherwise free
variables) and either effect free or wrapped in a frame to set the
allowed effects.  The testing rules will assume this.

Head labels and pointer labels alpha-vary and we consider machine
states, terms, etc., equal up to alpha-equivalence.  Scoping is as
follows: the $x$ in $\IFM{x}{t_1}{t_2}{\PH}{t_3}$ binds in $t_2$,
however, $\PH$ is a copy of the pointer heap and so the left-hand side
pointer labels are bound by the outer pointer heap and should
alpha-vary with those; in $\FRAME{\env}{t}{\fx}$ the variables in $\env$
bind in $t$ and the outer environment does not bind in $t$; in
$\clos{\env}{f}$ the variables in $\env$ bind in $f$ and the outer
environment does not bind in $f$; and in $\ms{\HH}{\PH}{\env}{\fx}{t}$
both $\HH$ and $\PH$ bind globally over the entire machine state and
$\env$ binds in $t$ except as noted above.

We write $\dom{\env}$, $\dom{\HH}$, and $\dom{\PH}$ to mean the set of
left-hand side variables, head labels, or pointer labels respectively.
When we write $\HH_1,\HH_2$ there is an implicit side condition that
$\dom{\HH_1}\cap\dom{\HH_2}=\litset{}$ and it means the obvious
concatenation, similarly for environments and pointer heaps.  We write
$\HH(\hl)$ to mean the $h$ that corresponds to $\hl$ in $\HH$ if there
is one.  When we write a predicate involving $\HH(\hl)$, we consider
the predicate to be false if $\hl\notin\dom{\HH}$.  Similarly for
environments and pointer heaps.

We write $\emptyseq$ to mean an empty sequence of definitions for an
environment, head heap, or pointer heap as the context dictates.

\section{Reduction Rules}

An abstraction machine state can take a step to a new state while
performing some action, it can fail, or it can commit a type error.
An action is either pure (no computational effects performd), the
input of an integer, the output of an integer, or a computational
effect (new, read, or write of a pointer).  Actions, ranged over by
metavariable $a$, are given by the grammar:
\[
\begin{array}{lcl}
a &\bnfdef& \AT \bnfalt \Ain{i} \bnfalt \Aout{i} \bnfalt \Anew \bnfalt \Ard \bnfalt \Awr \\
\end{array}
\]
The judgements for the reductions are:
\[
\begin{array}{l|l}
\hbox{Judgement} & \hbox{Meaning} \\\hline
M_1\stepsto{a}M_2 & \hbox{$M_1$ performs $a$ and reduces to $M_2$} \\
M\stepstoF & \hbox{$M$ fails} \\
M\stepstoE & \hbox{$M$ commits a type error} \\
\end{array}
\]

\subsection{Auxilary Predicates}

To express the rules concisely we use a few auxilary predicates on
heads:
\[
\begin{array}{lcl}
\Hisint{h} &=& \exists i.h=i \\
\Hisnat{h} &=& \exists i.h=i\land i\ge0 \\
\Histabf{i_1,\ldots,i_n}{h} &=& \exists \hl_1,\ldots,\hl_n.h=\TABLE{\tabeV{i_1}{\hl_1},\ldots,\tabeV{i_n}{\hl_n}} \\
\Histab{h} &=& \exists i_1,\ldots,i_n.\Histabf{i_1,\ldots,i_n}{h} \\
\Hisarr{h} &=& \exists \hl_1,\ldots,\hl_n.h=\TABLE{\tabeV{0}{\hl_1},\ldots,\tabeV{n-1}{\hl_n}} \\
\Hisfun{h} &=& \exists \env,f.h=\clos{\env}{f} \\
\Histyp{h} &=& \exists \env,x,t,\fx_1.h=\clos{\env}{\funL{x}{t}{\dkI}{\fx_1}{\fxT}{x}} \\
\Hisapp{h} &=& \Histab{h}\lor\Hisfun{h} \\
\Hisptr{h} &=& \exists \pl.h=\pl \\
\dkof{f} &=& \left\{\begin{array}{ll}
    \dk & f=\funL{x}{t_1}{\dk}{\fx_1}{\fx_2}{t_2} \\
    \dkC & f=\funA{x_1}{t_1}{t_2}{x_2}{t_3}{\fx_1}{\fx_2}{t_4} \\
  \end{array}\right. \\
\end{array}
\]

As previously mentioned, if $\hl\notin\dom\HH$ then
$\Hisint{\HH(\hl)}$ is false, and similarly for other predicates.

\subsection{Generate Mode}

\subsubsection{Contexts}

Various terms evaluate their subterms left to right first.  The follow rules
capture this behaviour uniformly.  Contexts are given by this grammar:
\[
\begin{array}{lcl}
E &\bnfdef&
  \uop\hole \bnfalt \bop{\hole}{t} \bnfalt \bop{\hl}{\hole} \bnfalt
  \cop{\hole}{t} \bnfalt \cop{\hl}{\hole} \bnfalt \arrL{\hole}{x}{t}
  \bnfalt {}\\&&
  \len\hole \bnfalt \appE{\hole}{t} \bnfalt \appE{\hl}{\hole}
  \bnfalt \appF{\hole}{t} \bnfalt \appF{\hl}{\hole}
  \bnfalt \ptrnew{t}{\hole} \bnfalt \ptrrd\hole \bnfalt \ptrwr{\hole}{t}
  \bnfalt \ptrwr{\hl}{\hole} \bnfalt \intout\hole \bnfalt {}\\&&
  \unify{\hole}{t} \bnfalt \LET{x}{\hole}{t} \\
  \end{array}
\]

\[
\infer[\rn{RGctxt}]
      {\ms{\HH_1}{\PH_1}{\env}{\fx}{\substhole{E}{t_1}}\stepsto{a}\ms{\HH_2}{\PH_2}{\env}{\fx}{\substhole{E}{t_2}}}
      {\ms{\HH_1}{\PH_1}{\env}{\fx}{t_1}\stepsto{a}\ms{\HH_2}{\PH_2}{\env}{\fx}{t_2}}
\]
\[
\infer[\rn{RGctxtF}]
      {\ms{\HH}{\PH}{\env}{\fx}{\substhole{E}{t_1}}\stepstoF}
      {\ms{\HH}{\PH}{\env}{\fx}{t_1}\stepstoF}
\]
\[
\infer[\rn{RGctxtE}]
      {\ms{\HH}{\PH}{\env}{\fx}{\substhole{E}{t_1}}\stepstoE}
      {\ms{\HH}{\PH}{\env}{\fx}{t_1}\stepstoE}
\]

\subsubsection{Variables}

A variable that is defined by the current environment steps to its
head label:
\[
\infer[\rn{RGvar}]
      {\ms{\HH}{\PH}{\env}{\fx}{x}\stepsto{\AT}\ms{\HH}{\PH}{\env}{\fx}{\hl}}
      {\env(x)=\hl}
\]

A variable that is not defined by the current environment commits an
error:
\[
\infer[\rn{RGvarE}]
      {\ms{\HH}{\PH}{\env}{\fx}{x}\stepstoE}
      {x\notin\dom{\env}}
\]

\subsubsection{None or All Values}

The uninhabited term fails:
\[
\infer[\rn{RGfalsesF}]
      {\ms{\HH}{\PH}{\env}{\fx}{\falses}\stepstoF}
      {}
\]

The all values term commits an error in generate mode:
\[
\infer[\rn{RGanysE}]
      {\ms{\HH}{\PH}{\env}{\fx}{\anys}\stepstoE}
      {}
\]

\subsubsection{Integers}

An integer steps to a new head label that is bound to that integer:
\[
\infer[\rn{RGi}]
      {\ms{\HH}{\PH}{\env}{\fx}{i}\stepsto{\AT}\ms{\HH,\hl=i}{\PH}{\env}{\fx}{\hl}}
      {}
\]

The all integer term commits an error in generate mode:
\[
\infer[\rn{RGintsE}]
      {\ms{\HH}{\PH}{\env}{\fx}{\ints}\stepstoE}
      {}
\]

\subsubsection{Integer Operations}

A unary operation applied to an integer reduces to the appropriate integer:
\[
\infer[\rn{RGuop}]
      {\ms{\HH}{\PH}{\env}{\fx}{\uop\hl}\stepsto{\AT}\ms{\HH}{\PH}{\env}{\fx}{i'}}
      {\HH(\hl)=i & i'=\uop{i}}
\]

A binary operation applied to integers reduces to the appropriate integer:
\[
\infer[\rn{RGbop}]
      {\ms{\HH}{\PH}{\env}{\fx}{\bop{\hl_1}{\hl_2}}\stepsto{\AT}\ms{\HH}{\PH}{\env}{\fx}{i}}
      {\HH(\hl_1)=i_1 & \HH(\hl_2)=i_2 & i=\bop{i_1}{i_2}}
\]

Comparisons applied to integers where the relationship holds reduce to
the first head label:
\[
\infer[\rn{RGcop}]
      {\ms{\HH}{\PH}{\env}{\fx}{\cop{\hl_1}{\hl_2}}\stepsto{\AT}\ms{\HH}{\PH}{\env}{\fx}{\hl_1}}
      {\HH(\hl_1)=i_1 & \HH(\hl_2)=i_2 & \cop{i_1}{i_2}}
\]

Comparisons applied to integers where the relationship does not hold fail:
\[
\infer[\rn{RGcopF}]
      {\ms{\HH}{\PH}{\env}{\fx}{\cop{\hl_1}{\hl_2}}\stepstoF}
      {\HH(\hl_1)=i_1 & \HH(\hl_2)=i_2 & \neg(\cop{i_1}{i_2})}
\]

Unary or binary operations or comparisons applied to non-integers commit errors:
\[
\infer[\rn{RGuopE}]
      {\ms{\HH}{\PH}{\env}{\fx}{\uop\hl}\stepstoE}
      {\neg(\Hisint{\HH(\hl)})}
\]
\[
\infer[\rn{RGbopE}]
      {\ms{\HH}{\PH}{\env}{\fx}{\bop{\hl_1}{\hl_2}}\stepstoE}
      {\neg(\Hisint{\HH(\hl_1)}\land\Hisint{\HH(\hl_2)})}
\]
\[
\infer[\rn{RGcopE}]
      {\ms{\HH}{\PH}{\env}{\fx}{\cop{\hl_1}{\hl_2}}\stepstoE}
      {\neg(\Hisint{\HH(\hl_1)}\land\Hisint{\HH(\hl_2)})}
\]

\subsubsection{Tables}

A fully evaluated fixed table steps to a new head label that is bound
to that table:
\[
\infer[\rn{RGtab1}]
      {\begin{array}{l}
         \ms{\HH}{\PH}{\env}{\fx}{\TABLE{\tabe{x_1}{i_1}{\hl_1},\ldots,\tabe{x_n}{i_n}{\hl_n}}}\stepsto{\AT} {}\\\quad
         \ms{\HH,\hl=\TABLE{\tabeV{i_1}{\hl_1},\ldots,\tabeV{i_n}{\hl_n}}}{\PH}{\env}{\fx}{\hl}
       \end{array}}
      {}
\]

Because fixed tables are dependent, we must specify the left-to-right
evaluation of subterms separately from the context rules above.
Specifically, a term gets evaluated in an environment that binds the
variables to the left to the already evaluated head labels:
\[
\infer[\rn{RGtab2}]
      {\ms{\HH_1}{\PH_1}{\env}{\fx} {t}\stepsto{a}\ms{\HH_2}{\PH_2}{\env}{\fx}{t'}}
      {\begin{array}{c}
         n\le m \\
         t = \TABLE{\tabe{x_1}{i_1}{\hl_1},\ldots,\tabe{x_{n-1}}{i_{n-1}}{\hl_{n-1}},\tabe{x_n}{i_n}{t_n},\ldots,
                    \tabe{x_m}{i_m}{t_m}} \\
         \env'=(\env,x_1=\hl_1,\ldots,x_{n-1}=\hl_{n-1}) \\
         \ms{\HH_1}{\PH_1}{\env'}{\fx}{t_n}\stepsto{a}\ms{\HH_2}{\PH_2}{\env'}{\fx}{t'_n} \\
         t' =
           \begin{array}[t]{l}
              \TABLE{\tabe{x_1}{i_1}{\hl_1},\ldots,\tabe{x_{n-1}}{i_{n-1}}{\hl_{n-1}},
                     \tabe{x_n}{i_n}{t'_n}, \\\phantom\langle
                     \tabe{x_{n+1}}{i_{n+1}}{t_{n+1}},\ldots,
                     \tabe{x_m}{i_m}{t_m}}
           \end{array} \\
       \end{array}}
\]
\[
\infer[\rn{RGtabF}]
      {\ms{\HH}{\PH}{\env}{\fx}{t}
       \stepstoF}
      {\begin{array}{c}
         n\le m \\
         t = \TABLE{\tabe{x_1}{i_1}{\hl_1},\ldots,\tabe{x_{n-1}}{i_{n-1}}{\hl_{n-1}},
                    \tabe{x_n}{i_n}{t_n},\ldots,\tabe{x_m}{i_m}{t_m}} \\
         \ms{\HH}{\PH}{\env,x_1=\hl_1,\ldots,x_{n-1}=\hl_{n-1}}{\fx}{t_n}\stepstoF \\
       \end{array}}
\]
\[
\infer[\rn{RGtabE}]
      {\ms{\HH}{\PH}{\env}{\fx}{t}
       \stepstoE}
      {\begin{array}{c}
         n\le m \\
         t = \TABLE{\tabe{x_1}{i_1}{\hl_1},\ldots,\tabe{x_{n-1}}{i_{n-1}}{\hl_{n-1}},
                    \tabe{x_n}{i_n}{t_n},\ldots,\tabe{x_m}{i_m}{t_m}} \\
         \ms{\HH}{\PH}{\env,x_1=\hl_1,\ldots,x_{n-1}=\hl_{n-1}}{\fx}{t_n}\stepstoE \\
       \end{array}}
\]

An array lambda applied to a natural number reduces to an appropriate fixed
table, and lets are used to bind the index variable appropriately:
\[
\infer[\rn{RGarr}]
      {\ms{\HH}{\PH}{\env}{\fx}{\arrL{\hl}{x}{t}}\stepsto{\AT}
       \ms{\HH}{\PH}{\env}{\fx}{t'}}
      {\begin{array}{c}
         \HH(\hl)=i \\
         i\ge0 \\
         y\notin\fv{t} \\
         t' = \TABLE{\tabe{y}{0}{\LET{x}{0}{t}},\ldots,\tabe{y}{i-1}{\LET{x}{i-1}{t}}} \\
       \end{array}}
\]

An array lambda applied to a non-natural number commits an error:
\[
\infer[\rn{RGarrE}]
      {\ms{\HH}{\PH}{\env}{\fx}{\arrL{\hl}{x}{t}}\stepstoE}
      {\neg(\Hisnat{\HH(\hl)})}
\]

The all tables term commits an error in generate mode:
\[
\infer[\rn{RGtabsE}]
      {\ms{\HH}{\PH}{\env}{\fx}{\tabs}\stepstoE}
      {}
\]

\subsubsection{Functions}

A function reduces to a closure using the current environment, so long
as it is contravariant or invaraint:
\[
\infer[\rn{RGfun}]
      {\ms{\HH}{\PH}{\env}{\fx}{f}\stepsto{\AT}\ms{\HH,\hl=\clos{\env}{f}}{\PH}{\env}{\fx}{\hl}}
      {\dkof{f}\in\litset{\dkC,\dkI}}
\]

A function that is not contravariant or invariant commits an error:
\[
\infer[\rn{RGfunE}]
      {\ms{\HH}{\PH}{\env}{\fx}{f}\stepstoE}
      {\dkof{f}\notin\litset{\dkC,\dkI}}
\]

The all functions term commits an error in generate mode:
\[
\infer[\rn{RGfunsE}]
      {\ms{\HH}{\PH}{\env}{\fx}{\funs}\stepstoE}
      {}
\]

\subsubsection{Length}

Length applied to an array reduces to the length of that array:
\[
\infer[\rn{RGlen}]
      {\ms{\HH}{\PH}{\env}{\fx}{\len\hl}\stepsto{\AT}\ms{\HH}{\PH}{\env}{\fx}{n}}
      {\HH(\hl)=\TABLE{\tabeV{0}{\hl_0},\ldots,\tabeV{n-1}{\hl_{n-1}}}}
\]

Length applied to a non-array commits an error:
\[
\infer[\rn{RGlenE}]
      {\ms{\HH}{\PH}{\env}{\fx}{\len\hl}\stepstoE}
      {\neg(\Hisarr{\HH(\hl)})}
\]

\subsubsection{Error Application}

A table applied to an integer that is in its domain reduces to the
corresponding element head label:
\[
\infer[\rn{RGappE1}]
      {\ms{\HH}{\PH}{\env}{\fx}{\appE{\hl_1}{\hl_2}}\stepsto{\AT}\ms{\HH}{\PH}{\env}{\fx}{\hl'_j}}
      {\HH(\hl_1)=\TABLE{\tabeV{i_1}{\hl'_1},\ldots,\tabeV{i_n}{\hl'_n}} &
       \HH(\hl_2)=i_j &
       1\le j\le n}
\]

A closure applied to a head label reduces to a frame to execute the
body in the closure's environment extended with the parameter bound to
the actual argument and to constrain the allowed effects to those
permitted by the effects annotation on the function:
\[
\infer[\rn{RGappE2}]
      {\ms{\HH}{\PH}{\env}{\fx}{\appE{\hl_1}{\hl_2}}\stepsto{\AT}
       \ms{\HH}{\PH}{\env}{\fx}{\FRAME{(\env',x=\hl_2)}{t_2}{\fx\fxmeet\fx_2}}}
      {\HH(\hl_1)=\clos{\env'}{\funL{x}{t_1}{\dk}{\fx_1}{\fx_2}{t_2}}}
\]
Note that we do not restrict the domain kind here as is done for
failing application.  We could make the rules more consistent, but
other rules prevent closures being created that are not contravariant
or invariant, so we keep the rule simple here.  In the case of an all
quantified contravariant function, we also run the all-quantifier
instantiation term to bind its variable:
\[
\infer[\rn{RGappE3}]
      {\ms{\HH}{\PH}{\env}{\fx}{\appE{\hl_1}{\hl_2}}\stepsto{\AT}
       \ms{\HH}{\PH}{\env}{\fx}{\FRAME{(\env',x_2=\hl_2)}{\LET{x_1}{t_2}{t_4}}{\fx\fxmeet\fx_2}}}
      {\HH(\hl_1)=\clos{\env'}{\funA{x_1}{t_1}{t_2}{x_2}{t_3}{\fx_1}{\fx_2}{t_4}}}
\]

A non-applicable applied to something commits an error:
\[
\infer[\rn{RGappEE1}]
      {\ms{\HH}{\PH}{\env}{\fx}{\appE{\hl_1}{\hl_2}}\stepstoE}
      {\neg(\Hisapp{\HH(\hl_1)})}
\]

A table applied to something that is not in its domain commits an
error:
\[
\infer[\rn{RGappEE2}]
      {\ms{\HH}{\PH}{\env}{\fx}{\appE{\hl_1}{\hl_2}}\stepstoE}
      {\HH(\hl_1)=\TABLE{\tabeV{i_1}{\hl'_1},\ldots,\tabeV{i_n}{\hl'_n}} &
       \neg(\HH(\hl_2)\in\litset{i_1,\ldots,i_n})}
\]

Note that it is an error to error apply a function outside its domain,
but as is standard in operational semantics, the runtime does not flag
this as an error, although the verifier should.

\subsubsection{Failing Application}

A table applied to an integer that is in its domain reduces to the
corresponding element head label:
\[
\infer[\rn{RGappF1}]
      {\ms{\HH}{\PH}{\env}{\fx}{\appF{\hl_1}{\hl_2}}\stepsto{\AT}\ms{\HH}{\PH}{\env}{\fx}{\hl'_j}}
      {\HH(\hl_1)=\TABLE{\tabeV{i_1}{\hl'_1},\ldots,\tabeV{i_n}{\hl'_n}} &
       \HH(\hl_2)=i_j &
       1\le j\le n}
\]

A table applied to something that is not in its domain fails (note
that $\hl_2$ must be defined in the head heap):
\[
\infer[\rn{RGappFF}]
      {\ms{\HH}{\PH}{\env}{\fx}{\appF{\hl_1}{\hl_2}}\stepstoF}
      {\HH(\hl_1)=\TABLE{\tabeV{i_1}{\hl'_1},\ldots,\tabeV{i_n}{\hl'_n}} &
       \HH(\hl_2)\notin\litset{i_1,\ldots,i_n}}
\]

The semantics of contravariant functions is such that even failure
applying one outside its domain is an error.  As already mentioned,
the runtime does not flag this, but the verifier should.  Thus a
closure with a contravariant function applied to a head label reduces
to a frame to execute the body in the closure's environment extended
with the parameter bound to the actual argument and to constrain the
allowed effects to those permitted by the effects annotation on the
function:
\[
\infer[\rn{RGappF2}]
      {\ms{\HH}{\PH}{\env}{\fx}{\appF{\hl_1}{\hl_2}}\stepsto{\AT}
       \ms{\HH}{\PH}{\env}{\fx}{\FRAME{(\env',x=\hl_2)}{t_2}{\fx\fxmeet\fx_2}}}
      {\HH(\hl_1)=\clos{\env'}{\funL{x}{t_1}{\dkC}{\fx_1}{\fx_2}{t_2}}}
\]
In the case of an all quantified contravariant function, we also run
the all-quantifier instantiation term to bind its variable:
\[
\infer[\rn{RGappF3}]
      {\ms{\HH}{\PH}{\env}{\fx}{\appF{\hl_1}{\hl_2}}\stepsto{\AT}
       \ms{\HH}{\PH}{\env}{\fx}{\FRAME{(\env',x_2=\hl_2)}{\LET{x_1}{t_2}{t_4}}{\fx\fxmeet\fx_2}}}
      {\HH(\hl_1)=\clos{\env'}{\funA{x_1}{t_1}{t_2}{x_2}{t_3}{\fx_1}{\fx_2}{t_4}}}
\]

In contrast failure applying an invariant function outside its domain
does actually fail.  Thus a closure with an invariant function applied
to a head label reduces to a test to check the actual argument is in
the domain, the domain term is wrapped in a frame using the closure's
environment and constraining the allowed effects according to the
effects annotation, if the test succeeds a frame is used to execute
the body in the closure's environment extended with the parameter
bound to the actual argument and to constrain the allowed effects to
those permitted by the effects annotation on the function:
\[
\infer[\rn{RGappF4}]
      {\begin{array}{l}
         \ms{\HH}{\PH}{\env}{\fx}{\appF{\hl_1}{\hl_2}}\stepsto{\AT} {}\\\quad
         \ms{\HH}{\PH}{\env}{\fx}
            {\FRAME{\env'}{\chk{\hl_2}{\emptyset}{t_1}{\FRAME{(\env',x=\hl_2)}{t_2}{\fx\fxmeet\fx_2}}{\falses}}
                   {\fx\fxmeet\fx_1}}
       \end{array}}
      {\HH(\hl_1)=\clos{\env'}{\funL{x}{t_1}{\dkI}{\fx_1}{\fx_2}{t_2}}}
\]

A non-applicable applied to something commits an error:
\[
\infer[\rn{RGappFE1}]
      {\ms{\HH}{\PH}{\env}{\fx}{\appF{\hl_1}{\hl_2}}\stepstoE}
      {\neg(\Hisapp{\HH(\hl_1)})}
\]

Applying a table to something not defined in the head heap commits an error
\[
\infer[\rn{RGappFE2}]
      {\ms{\HH}{\PH}{\env}{\fx}{\appF{\hl_1}{\hl_2}}\stepstoE}
      {\HH(\hl_1)=\TABLE{\tabeV{i_1}{\hl'_1},\ldots,\tabeV{i_n}{\hl'_n}} & \hl_2\notin\dom{\HH}}
\]

Applying a function that is not contravariant or invariant commits an
error:
\[
\infer[\rn{RGappFE3}]
      {\ms{\HH}{\PH}{\env}{\fx}{\appF{\hl_1}{\hl_2}}\stepstoE}
      {\HH(\hl_1)=\clos{\env'}{f} & \dkof{f}\notin\litset{\dkC,\dkI}}
\]

\subsubsection{From}

From commits an error in generate mode:
\[
\infer[\rn{RGfromE}]
      {\ms{\HH}{\PH}{\env}{\fx}{\from{t}}\stepstoE}
      {}
\]

\subsubsection{Pointers}

A pointer creation reduces to a new head label bound to a new pointer
label, which is initially bound to the initial head label, so long as
new effects are allowed:
\[
\infer[\rn{RGnew}]
      {\ms{\HH}{\PH}{\env}{\fx}{\ptrnew{t}{\hl}}\stepsto{\Anew}
       \ms{\HH,\hl'=\pl}{\PH,\pl=\ptrc{\env}{t}{\hl}}{\env}{\fx}{\hl'}}
      {\litset{\fxN}\fxle\fx}
\]

A read of a pointer reduces to its current contents, so long as read
effects are allowed:
\[
\infer[\rn{RGread}]
      {\ms{\HH}{\PH}{\env}{\fx}{\ptrrd\hl}\stepsto{\Ard}\ms{\HH}{\PH}{\env}{\fx}{\hl'}}
      {\HH(\hl)=\pl & \PH(\pl)=\ptrc{\env'}{t'}{\hl'} & \litset{\fxR}\fxle\fx}
\]

A write of a head label to a pointer, updates the contents of the
pointer and reduces to the head label, so long as write effects are
allowed:
\[
\infer[\rn{RGwrite}]
      {\ms{\HH}{\PH,\pl=\ptrc{\env'}{t'}{\hl'}}{\env}{\fx}{\ptrwr{\hl_1}{\hl_2}}\stepsto{\Awr}
       \ms{\HH}{\PH,\pl=\ptrc{\env'}{t'}{\hl_2}}{\env}{\fx}{\hl_2}}
      {\HH(\hl_1)=\pl & \litset{\fxW}\fxle\fx}
\]

A pointer creation when new effects are not allowed commits an error:
\[
\infer[\rn{RGnewE}]
      {\ms{\HH}{\PH}{\env}{\fx}{\ptrnew{t}{\hl}}\stepstoE}
      {\litset{\fxN}\not\fxle\fx}
\]

A read or write of a non-pointer or a pointer not defined in the
pointer heap commits an error, as do those operation when the
respective effects are not allowed:
\[
\infer[\rn{RGreadE}]
      {\ms{\HH}{\PH}{\env}{\fx}{\ptrrd{\hl}}\stepstoE}
      {\neg(\Hisptr{\HH(\hl)})\lor(\HH(\hl)=\pl\land\pl\notin\dom\PH)\lor\litset{\fxR}\not\fxle\fx}
\]
\[
\infer[\rn{RGwriteE}]
      {\ms{\HH}{\PH}{\env}{\fx}{\ptrwr{\hl_1}{\hl_2}}\stepstoE}
      {\neg(\Hisptr{\HH(\hl_1)})\lor(\HH(\hl_1)=\pl\land\pl\notin\dom\PH)\lor\litset{\fxW}\not\fxle\fx}
\]

The pointer type term commits an error in both generate and test mode,
and the all pointers term commits an error in generate mode:
\[
\infer[\rn{RGptrE}]
      {\ms{\HH}{\PH}{\env}{\fx}{\ptrto{t}}\stepstoE}
      {}
\]
\[
\infer[\rn{RGptrsE}]
      {\ms{\HH}{\PH}{\env}{\fx}{\ptrs}\stepstoE}
      {}
\]

\subsubsection{Input and Output}

Input reduces to the integer inputted, which appears in the action, so
long as IO effects are allowed:
\[
\infer[\rn{RGin}]
      {\ms{\HH}{\PH}{\env}{\fx}{\intin}\stepsto{\Ain{i}}\ms{\HH}{\PH}{\env}{\fx}{i}}
      {\litset{\fxIO}\fxle\fx}
\]

Output of an integer reduces to the same head label and has the output
of that integer as the action, so long as IO effects are allowed:
\[
\infer[\rn{RGout}]
      {\ms{\HH}{\PH}{\env}{\fx}{\intout{\hl}}\stepsto{\Aout{i}}\ms{\HH}{\PH}{\env}{\fx}{\hl}}
      {\HH(\hl)=i & \litset{\fxIO}\fxle\fx}
\]

Input when IO effects are not allowed commits an error:
\[
\infer[\rn{RGinE}]
      {\ms{\HH}{\PH}{\env}{\fx}{\intin}\stepstoE}
      {\litset{\fxIO}\not\fxle\fx}
\]

Output of a non-integer commits an error, as does output when IO
effects are not allowed:
\[
\infer[\rn{RGoutE}]
      {\ms{\HH}{\PH}{\env}{\fx}{\intout{\hl}}\stepstoE}
      {\neg(\Hisint{\HH(\hl)})\lor\litset{\fxIO}\not\fxle\fx}
\]

\subsubsection{Unify and Join}

Unification reduces to a test against right-hand term, which if it
succeeds produces the head label:
\[
\infer[\rn{RGunify}]
      {\ms{\HH}{\PH}{\env}{\fx}{\unify{\hl}{t}}\stepsto{\AT}
       \ms{\HH}{\PH}{\env}{\fx}{\chk{\hl}{\emptyset}{t}{\hl}{\falses}}}
      {}
\]

Join commits an error in generate mode:
\[
\infer[\rn{RGjoinE}]
      {\ms{\HH}{\PH}{\env}{\fx}{\join{t_1}{t_2}}\stepstoE}
      {}
\]

\subsubsection{Let}

Let reduces to a frame:
\[
\infer[\rn{RGlet}]
      {\ms{\HH}{\PH}{\env}{\fx}{\LET{x}{\hl}{t}}\stepsto{\AT}\ms{\HH}{\PH}{\env}{\fx}{\FRAME{(\env,x=\hl)}{t}{\fx}}}
      {}
\]

\subsubsection{Letrec}

The rules for letrec require two auxilary judgements for evaluating
values to heads or errors.  Judgement $\evalV{\PH_1}{\env}{v}{h}{\PH_2}$
means that value $v$ evaluates to head $h$ in environment $\env$,
possible extending the pointer heap from $\PH_1$ to $\PH_2$.  Judgement
$\evalVE{\env}{v}$ means that value $v$ is erroneous in environment
$\env$.

To evaluate a letrec, we allocate a new head label for each variable
and form the new environment, using that environment, we evaluate each
value to a head, possibly extending the pointer heap with new
pointers, form the new head heap, and then execute the body in the new
environment, head heap, and pointer heap.  Additionally, if any of the
values create pointers, new effects must be allowed, and the action
reflects new effects.
\[
\infer[\rn{RGletrec}]
      {\ms{\HH}{\PH_0}{\env}{\fx}{\LETREC{x_1=v_1,\ldots,x_n=v_n}{t}}\stepsto{a}
       \ms{\HH'}{\PH_n}{\env}{\fx}{\FRAME{\env'}{t}{\fx}}}
      {\begin{array}{c}
         \env'=(\env,x_1=\hl_1,\ldots,x_n=\hl_n) \\
         \evalV{\PH_0}{\env'}{v_1}{h_1}{\PH_1} \\
         \vdots \\
         \evalV{\PH_{n-1}}{\env'}{v_n}{h_n}{\PH_n} \\
         \HH' = (\HH,\hl_1=h_1,\ldots,\hl_n=h_n) \\
         {\it ptrs} = \exists j,t,x:1\le j\le n\land v_j=\ptrnew{t}{x} \\
         {\it ptrs} \implies \litset{\fxN}\fxle\fx \\
         a = \left\{\begin{array}{ll} \Anew & {\it ptrs} \\ \AT & \neg{\it ptrs} \end{array}\right. \\
       \end{array}}
\]

\[
\infer[\rn{RVtable}]
      {\evalV{\PH}{\env}{\TABLE{\tabeV{i_1}{x_1},\ldots,\tabeV{i_n}{x_n}}}
                        {\TABLE{\tabeV{i_1}{\hl_1},\ldots,\tabeV{i_n}{\hl_n}}}{\PH}}
      {\env(x_1)=\hl_1 & \cdots & \env(x_n)=\hl_n}
\]
\[
\infer[\rn{RVfun}]
      {\evalV{\PH}{\env}{f}{\clos{\env}{f}}{\PH}}
      {\dkof{f}\in\litset{\dkC,\dkI}}
\]
\[
\infer[\rn{RVptr}]
      {\evalV{\PH}{\env}{\ptrnew{t}{x}}{\pl}{\PH,\pl=\ptrc{\env}{t}{\hl}}}
      {\env(x)=\hl}
\]

If any value is errorneous then letrec commits an error:
\[
\infer[\rn{RGletrecE1}]
      {\ms{\HH}{\PH}{\env}{\fx}{\LETREC{x_1=v_1,\ldots,x_n=v_n}{t}}\stepstoE}
      {\env'=(\env,x_1=\hl_1,\ldots,x_n=\hl_n) & 1\le j\le n & \evalVE{\env'}{v_j}}
\]
Note that head labels used in this rule can be arbitrary---they are
not relevant to a value being erroneous.

If any value is a pointer and new effects are not allowed then letrec
commits an error:
\[
\infer[\rn{RGletrecE2}]
      {\ms{\HH}{\PH}{\env}{\fx}{\LETREC{x_1=v_1,\ldots,x_n=v_n}{t}}\stepstoE}
      {1\le j\le n & v_j=\ptrnew{t}{x} & \litset{\fxN}\not\fxle\fx}
\]

A tuple is errorenous if one of the variables is out
of scope:
\[
\infer[\rn{RVtableE}]
      {\evalVE{\env}{\TABLE{\tabeV{i_1}{x_1},\ldots,\tabeV{i_n}{x_n}}}}
      {1\le j\le n & x_j\notin\dom{\env}}
\]

A function value is erroneous if it is not contravariant or invariant:
\[
\infer[\rn{RVfunE}]
      {\evalVE{\env}{f}}
      {\dkof{f}\notin\litset{\dkC,\dkI}}
\]

A pointer value is errorenous if the variable is out of scope:
\[
\infer[\rn{RVptrE}]
      {\evalVE{\env}{\ptrnew{t}{x}}}
      {x\notin\dom{\env}}
\]

\subsubsection{Conditionals}

A conditional saves the current pointer heap (so it can be restored if
the condition fails) and then executes according to the remaining
rules in this section:
\[
\infer[\rn{RGif}]
      {\ms{\HH}{\PH}{\env}{\fx}{\IF{x}{t_1}{t_2}{t_3}}\stepsto{\AT}
       \ms{\HH}{\PH}{\env}{\fx}{\IFM{x}{t_1}{t_2}{\PH}{t_3}}}
      {}
\]

If the condition is fully evaluated then a conditional reduces to a frame:
\[
\infer[\rn{RGif1}]
      {\ms{\HH}{\PH}{\env}{\fx}{\IFM{x}{\hl}{t_2}{\PH'}{t_3}}\stepsto{\AT}
       \ms{\HH}{\PH}{\env}{\fx}{\FRAME{(\env,x=\hl)}{t_2}{\fx}}}
      {}
\]

If the condition takes a step then the conditional takes a step, note
the restriction of effects to irreversible ones:
\[
\infer[\rn{RGif2}]
      {\ms{\HH_1}{\PH_1}{\env}{\fx}{\IFM{x}{t_1}{t_2}{\PH'}{t_3}}\stepsto{a}
       \ms{\HH_2}{\PH_2}{\env}{\fx}{\IFM{x}{t'_1}{t_2}{\PH'}{t_3}}}
      {\fx'=\fx\fxmeet\fxRV & \ms{\HH_1}{\PH_1}{\env}{\fx'}{t_1}\stepsto{a}\ms{\HH_2}{\PH_2}{\env}{\fx'}{t'_1}}
\]

If the condition fails then the conditional reduces to the false branch
after restoring the pointer heap:
\[
\infer[\rn{RGif3}]
      {\ms{\HH}{\PH}{\env}{\fx}{\IFM{x}{t_1}{t_2}{\PH'}{t_3}}\stepsto{\AT}
       \ms{\HH}{\subst{\PH}{\PH'}}{\env}{\fx}{t_3}}
      {\ms{\HH}{\PH}{\env}{\fx\fxmeet\fxRV}{t_1}\stepstoF}
\]
where $\subst{\PH}{\PH'}$ is $\PH$ but with the contents of each
pointer replaced by its contents in $\PH'$ where defined.  We could
use the simpler rule of making the pointer heap in the reduced machine
state be $\PH'$, but that makes it harder to state an invariant needed
to prove type safey, so we use this more complicated rule that keeps
around dead pointer labels.

If the condition commits an error then so does the conditional:
\[
\infer[\rn{RGifE}]
      {\ms{\HH}{\PH}{\env}{\fx}{\IFM{x}{\hl}{t_2}{\PH'}{t_3}}\stepstoE}
      {\ms{\HH}{\PH}{\env}{\fx\fxmeet\fxRV}{t_1}\stepstoE}
\]

\subsubsection{Stage}

A stage reduces to the right-hand term:
\[
\infer[\rn{RGstage}]
      {\ms{\HH}{\PH}{\env}{\fx}{\stage{\fx'}{d}{t_1}{t_2}}\stepsto{\AT}\ms{\HH}{\PH}{\env}{\fx}{t_2}}
      {}
\]
Note that the effects and decidability annotations are not checked by
the runtime.

\subsubsection{Effects Then}

An effects then term commits an error:
\[
\infer[\rn{RGfxE}]
      {\ms{\HH}{\PH}{\env}{\fx}{\fxthen{\fx'}{t}}\stepstoE}
      {}
\]

\subsubsection{Frames}

Frames just change the environment and allowed effects:
\[
\infer[\rn{RGframe1}]
      {\ms{\HH}{\PH}{\env}{\fx}{\FRAME{\env'}{\hl}{\fx'}}\stepsto{\AT}\ms{\HH}{\PH}{\env}{\fx}{\hl}}
      {}
\]
\[
\infer[\rn{RGframe2}]
      {\ms{\HH_1}{\PH_1}{\env}{\fx}{\FRAME{\env'}{t_1}{\fx'}}\stepsto{a}
       \ms{\HH_2}{\PH_2}{\env}{\fx}{\FRAME{\env'}{t_2}{\fx'}}}
      {\ms{\HH_1}{\PH_1}{\env'}{\fx'}{t_1}\stepsto{a}\ms{\HH_2}{\PH_2}{\env'}{\fx'}{t_2}}
\]
\[
\infer[\rn{RGframeF}]
      {\ms{\HH}{\PH}{\env}{\fx}{\FRAME{\env'}{t}{\fx'}}\stepstoF}
      {\ms{\HH}{\PH}{\env'}{\fx'}{t}\stepstoF}
\]
\[
\infer[\rn{RGframeE}]
      {\ms{\HH}{\PH}{\env}{\fx}{\FRAME{\env'}{t}{\fx'}}\stepstoE}
      {\ms{\HH}{\PH}{\env'}{\fx'}{t}\stepstoE}
\]

\subsection{Test Mode}

\subsubsection{Revert to Generate Mode}

Certain terms are tested against simply by generating their values
and then comparing the values.  This is captured by the following
uniform rule.  The revert to generate terms are given by the following
grammar:
\[
\begin{array}{lcl}
g &\bnfdef&
  \uop{t} \bnfalt \bop{t_1}{t_2} \bnfalt \len{t} \bnfalt
  \appE{t_1}{t_2} \bnfalt \appF{t_1}{t_2} \bnfalt \ptrnew{t_1}{t_2}
  \bnfalt \ptrrd{t} \bnfalt \ptrwr{t_1}{t_2} \bnfalt \ptrto{t} \bnfalt
  \intin \bnfalt \intout{t} \bnfalt \fxthen{\fx}{t} \\
\end{array}
\]

\[
\infer[\rn{RTgen}]
      {\ms{\HH}{\PH}{\env}{\fx}{\chk{\hl}{\AEQ}{g}{t_1}{t_2}}\stepsto{\AT}
       \ms{\HH}{\PH}{\env}{\fx}{\LET{x}{g}{\chk{\hl}{\AEQ}{x}{t_1}{t_2}}}}
      {}
\]

\subsubsection{Variables}

Testing against a variable reduces to testing against the head label
it is bound to:
\[
\infer[\rn{RTvar}]
      {\ms{\HH}{\PH}{\env}{\fx}{\chk{\hl}{\AEQ}{x}{t_1}{t_2}}\stepsto{\AT}
       \ms{\HH}{\PH}{\env}{\fx}{\chk{\hl}{\AEQ}{\hl'}{t_1}{t_2}}}
      {\env(x)=\hl'}
\]

Testing against an out of scope variable commits an error:
\[
\infer[\rn{RTvarE}]
      {\ms{\HH}{\PH}{\env}{\fx}{\chk{\hl}{\AEQ}{x}{t_1}{t_2}}\stepstoE}
      {x\notin\dom\env}
\]

\subsubsection{None or All Values}

Testing against no values reduces to the false branch:
\[
\infer[\rn{RTfalses}]
      {\ms{\HH}{\PH}{\env}{\fx}{\chk{\hl}{\AEQ}{\falses}{t_1}{t_2}}\stepsto{\AT}
       \ms{\HH}{\PH}{\env}{\fx}{t_2}}
      {}
\]

Testing against all values reduces to the true branch:
\[
\infer[\rn{RTanys}]
      {\ms{\HH}{\PH}{\env}{\fx}{\chk{\hl}{\AEQ}{\anys}{t_1}{t_2}}\stepsto{\AT}
       \ms{\HH}{\PH}{\env}{\fx}{t_1}}
      {}
\]

\subsubsection{Integers}

Testing against a specific integer, checks for that integer:
\[
\infer[\rn{RTi1}]
      {\ms{\HH}{\PH}{\env}{\fx}{\chk{\hl}{\AEQ}{i}{t_1}{t_2}}\stepsto{\AT}
       \ms{\HH}{\PH}{\env}{\fx}{t_1}}
      {\HH(\hl)=i}
\]
\[
\infer[\rn{RTi2}]
      {\ms{\HH}{\PH}{\env}{\fx}{\chk{\hl}{\AEQ}{i}{t_1}{t_2}}\stepsto{\AT}
       \ms{\HH}{\PH}{\env}{\fx}{t_2}}
      {\HH(\hl)\ne i}
\]

Testing against all integers, checks for integers:
\[
\infer[\rn{RTints1}]
      {\ms{\HH}{\PH}{\env}{\fx}{\chk{\hl}{\AEQ}{\ints}{t_1}{t_2}}\stepsto{\AT}
       \ms{\HH}{\PH}{\env}{\fx}{t_1}}
      {\Hisint{\HH(\hl)}}
\]
\[
\infer[\rn{RTints2}]
      {\ms{\HH}{\PH}{\env}{\fx}{\chk{\hl}{\AEQ}{\ints}{t_1}{t_2}}\stepsto{\AT}
       \ms{\HH}{\PH}{\env}{\fx}{t_2}}
      {\hl\in\dom\HH & \neg\Hisint{\HH(\hl)}}
\]

Testing an undefined head label against integers or all integers
commits an error:
\[
\infer[\rn{RTiE}]
      {\ms{\HH}{\PH}{\env}{\fx}{\chk{\hl}{\AEQ}{i}{t_1}{t_2}}\stepstoE}
      {\hl\notin\dom\HH}
\]
\[
\infer[\rn{RTintsE}]
      {\ms{\HH}{\PH}{\env}{\fx}{\chk{\hl}{\AEQ}{\ints}{t_1}{t_2}}\stepstoE}
      {\hl\notin\dom\HH}
\]

\subsubsection{Integer Comparisons}

To test against a comparison, first test against the left subterm,
then compute the second term, then do the comparison:
\[
\infer[\rn{RTcop}]
      {\begin{array}{l}
         \ms{\HH}{\PH}{\env}{\fx}{\chk{\hl}{\AEQ}{\cop{t'_1}{t'_2}}{t_1}{t_2}}\stepsto{\AT} {}\\\quad
       \ms{\HH}{\PH}{\env}{\fx}{\chk{\hl}{\AEQ}{t'_1}{\IF{x}{\cop{\hl}{\FRAME{\env}{t'_2}{\fx}}}{t_1}{t_2}}{t_2}}
       \end{array}}
      {}
\]
Note we use a frame here to keep things closed and the correct allowed
effects, as required by the syntax.

\subsubsection{Tables}

To test against a fixed table, the head label must map to a fixed
table with the same domain, then each label is tested against the
corresponding term, in sequence, binding the variables as well:
\[
\infer[\rn{RTtab1}]
      {\begin{array}{l}
         \ms{\HH}{\PH}{\env}{\fx}
            {\chk{\hl}{\AEQ}{\TABLE{\tabe{x_1}{i_1}{t_1},\ldots,\tabe{x_n}{i_n}{t_n}}}{t'_1}{t'_2}}\stepsto{\AT}
         {}\\\quad
         \ms{\HH}{\PH}{\env}{\fx}{\chk{\hl_1}{\AEQ}{t_1}{\FRAME{(\env,x_1=\hl_1)}
                                      {\cdots\chk{\hl_n}{\AEQ}{t_n}{t'_1}{t'_2}\cdots}{\fx}}{t'_2}}
       \end{array}}
      {\HH(\hl)=\TABLE{\tabeV{i_1}{\hl_1},\ldots,\tabeV{i_n}{\hl_n}}}
\]
\[
\infer[\rn{RTtab2}]
      {\ms{\HH}{\PH}{\env}{\fx}
          {\chk{\hl}{\AEQ}{\TABLE{\tabe{x_1}{i_1}{t_1},\ldots,\tabe{x_n}{i_n}{t_n}}}{t'_1}{t'_2}}\stepsto{\AT}
       \ms{\HH}{\PH}{\env}{\fx}{t'_2}}
      {\hl\in\dom{\HH} & \neg(\Histabf{i_1,\ldots,i_n}{\HH(\hl)})}
\]

To test against an array lambda, the head label must map to an array,
then the length is tested against the length term, then each head
label in the array is tested against the range term with the index
variable bound to the index:
\[
\infer[\rn{RTarr1}]
      {\ms{\HH}{\PH}{\env}{\fx}{\chk{\hl}{\AEQ}{\arrL{t_1}{x}{t_2}}{t'_1}{t'_2}}\stepsto{\AT}
       \ms{\HH,\hl'=n}{\PH}{\env}{\fx}{t}}
      {\begin{array}{c}
         \HH(\hl)=\TABLE{\tabeV{0}{\hl_1},\ldots,\tabeV{n-1}{\hl_n}} \\
         t=\begin{array}[t]{l}
           \chk{\hl'}{\AEQ}{t_1}{\\
           \FRAME{\env}{\chk{\hl_1}{\AEQ}{\LET{x}{0}{t_2}}{\\
           \vdots \\
           \FRAME{\env}{\chk{\hl_n}{\AEQ}{\LET{x}{n-1}{t_2}}{\\
           t'_1}{t'_2}}{\fx}\cdots}{t'_2}}{\fx}}{t'_2}
         \end{array}
       \end{array}}
\]
\[
\infer[\rn{RTarr2}]
      {\ms{\HH}{\PH}{\env}{\fx}{\chk{\hl}{\AEQ}{\arrL{t_1}{x}{t_2}}{t'_1}{t'_2}}\stepsto{\AT}
       \ms{\HH}{\PH}{\env}{\fx}{t'_2}}
      {\hl\in\dom{\HH} & \neg(\Hisarr{\HH(\hl)})}
\]

Testing against all tables, checks for tables:
\[
\infer[\rn{RTtabs1}]
      {\ms{\HH}{\PH}{\env}{\fx}{\chk{\hl}{\AEQ}{\tabs}{t_1}{t_2}}\stepsto{\AT}
       \ms{\HH}{\PH}{\env}{\fx}{t_1}}
      {\Histab{\HH(\hl)}}
\]
\[
\infer[\rn{RTtabs2}]
      {\ms{\HH}{\PH}{\env}{\fx}{\chk{\hl}{\AEQ}{\tabs}{t_1}{t_2}}\stepsto{\AT}
       \ms{\HH}{\PH}{\env}{\fx}{t_2}}
      {\hl\in\dom\HH & \neg\Histab{\HH(\hl)}}
\]

Testing an undefined head label against a fixed table, an array
lambda, or all tables commits an error:
\[
\infer[\rn{RTtabE}]
      {\ms{\HH}{\PH}{\env}{\fx}
          {\chk{\hl}{\AEQ}{\TABLE{\tabe{x_1}{i_1}{t_1},\ldots,\tabe{x_n}{i_n}{t_n}}}{t'_1}{t'_2}}\stepstoE}
      {\hl\notin\dom\HH}
\]
\[
\infer[\rn{RTarrE}]
      {\ms{\HH}{\PH}{\env}{\fx}{\chk{\hl}{\AEQ}{\arrL{t_1}{x}{t_2}}{t'_1}{t'_2}}\stepstoE}
      {\hl\notin\dom\HH}
\]
\[
\infer[\rn{RTtabsE}]
      {\ms{\HH}{\PH}{\env}{\fx}{\chk{\hl}{\AEQ}{\tabs}{t_1}{t_2}}\stepstoE}
      {\hl\notin\dom\HH}
\]

\subsubsection{Functions}

Testing a non-function against a function fails:
\[
\infer[\rn{RTfun}]
      {\ms{\HH}{\PH}{\env}{\fx}{\chk{\hl}{\AEQ}{f}{t_1}{t_2}}\stepsto{\AT}
       \ms{\HH}{\PH}{\env}{\fx}{t_2}}
      {\hl\in\dom\HH & \neg(\Hisfun{\HH(\hl)})}
\]

Testing against all functions, checks for closures:
\[
\infer[\rn{RTfuns1}]
      {\ms{\HH}{\PH}{\env}{\fx}{\chk{\hl}{\AEQ}{\funs}{t_1}{t_2}}\stepsto{\AT}
       \ms{\HH}{\PH}{\env}{\fx}{t_1}}
      {\Hisfun{\HH(\hl)}}
\]
\[
\infer[\rn{RTfuns2}]
      {\ms{\HH}{\PH}{\env}{\fx}{\chk{\hl}{\AEQ}{\funs}{t_1}{t_2}}\stepsto{\AT}
       \ms{\HH}{\PH}{\env}{\fx}{t_2}}
      {\hl\in\dom\HH & \neg(\Hisfun{\HH(\hl)})}
\]

Testing a function against another function commits an error:
\[
\infer[\rn{RTfunE1}]
      {\ms{\HH}{\PH}{\env}{\fx}{\chk{\hl}{\AEQ}{f}{t_1}{t_2}}\stepstoE}
      {\HH(\hl)=\clos{\env'}{f'}}
\]

Testing an undefined head label against a function or all functions
commits an error:
\[
\infer[\rn{RTfunE2}]
      {\ms{\HH}{\PH}{\env}{\fx}{\chk{\hl}{\AEQ}{f}{t_1}{t_2}}\stepstoE}
      {\hl\notin\dom\HH}
\]
\[
\infer[\rn{RTfunsE}]
      {\ms{\HH}{\PH}{\env}{\fx}{\chk{\hl}{\AEQ}{\funs}{t_1}{t_2}}\stepstoE}
      {\hl\notin\dom\HH}
\]

\subsubsection{From}

To test against a from, first evaluate the subterm to an invariant
syntactic identity function, and then test against the domain term in
a frame based on the closure:
\[
\infer[\rn{RTfrom1}]
      {\ms{\HH}{\PH}{\env}{\fx}{\chk{\hl}{\AEQ}{\from{t}}{t_1}{t_2}}\stepsto{\AT}
       \ms{\HH}{\PH}{\env}{\fx}{\LET{x}{t}{\chk{\hl}{\AEQ}{\from{x}}{t_1}{t_2}}}}
      {\forall y.t\ne y}
\]
\[
\infer[\rn{RTfrom2}]
      {\ms{\HH}{\PH}{\env}{\fx}{\chk{\hl}{\AEQ}{\from{x}}{t_1}{t_2}}\stepsto{\AT}
       \ms{\HH}{\PH}{\env}{\fx}{\FRAME{\env'}{\chk{\hl}{\AEQ}{t'_1}{t_1}{t_2}}{\fx\fxmeet\fx_1}}}
      {\env(x)=\hl' & \HH(\hl')=\clos{\env'}{\funL{x}{{t'_1}}{\dkI}{\fx_1}{\fxT}{x}}}
\]

To test against a non-invariant syntactic function or against an
undefined function commits an error:
\[
\infer[\rn{RTfromE}]
      {\ms{\HH}{\PH}{\env}{\fx}{\chk{\hl}{\AEQ}{\from{x}}{t_1}{t_2}}\stepstoE}
      {x\notin\dom{\env}\lor(\env(x)=\hl'\land\neg(\Histyp{\HH(\hl')}))}
\]

\subsubsection{Pointers}

Testing against all pointers, checks for pointers:
\[
\infer[\rn{RTptrs1}]
      {\ms{\HH}{\PH}{\env}{\fx}{\chk{\hl}{\AEQ}{\ptrs}{t_1}{t_2}}\stepsto{\AT}
       \ms{\HH}{\PH}{\env}{\fx}{t_1}}
      {\Hisptr{\HH(\hl)}}
\]
\[
\infer[\rn{RTptrs2}]
      {\ms{\HH}{\PH}{\env}{\fx}{\chk{\hl}{\AEQ}{\ptrs}{t_1}{t_2}}\stepsto{\AT}
       \ms{\HH}{\PH}{\env}{\fx}{t_2}}
      {\hl\in\dom\HH & \neg\Hisptr{\HH(\hl)}}
\]

Testing an undefined head label against all pointers
commits an error:
\[
\infer[\rn{RTptrsE}]
      {\ms{\HH}{\PH}{\env}{\fx}{\chk{\hl}{\AEQ}{\ptrs}{t_1}{t_2}}\stepstoE}
      {\hl\notin\dom\HH}
\]

\subsubsection{Unify}

To test against a unify, test against each subterm:
\[
\infer[\rn{RTunify}]
      {\begin{array}{l}
         \ms{\HH}{\PH}{\env}{\fx}{\chk{\hl}{\AEQ}{\unify{t'_1}{t'_2}}{t_1}{t_2}}\stepsto{\AT} {}\\\quad
         \ms{\HH}{\PH}{\env}{\fx}{\chk{\hl}{\AEQ}{t'_1}{\FRAME{\env}{\chk{\hl}{\AEQ}{t'_2}{t_1}{t_2}}{\fx}}{t_2}}
       \end{array}}
      {}
\]

\subsubsection{Join}

To test against a join, test against the first subterm and if that
fails test against the second subterm:
\[
\infer[\rn{RTjoin}]
      {\begin{array}{l}
         \ms{\HH}{\PH}{\env}{\fx}{\chk{\hl}{\AEQ}{\join{t'_1}{t'_2}}{t_1}{t_2}}\stepsto{\AT} {}\\\quad
         \ms{\HH}{\PH}{\env}{\fx}{\chk{\hl}{\AEQ}{t'_1}{t_1}{\FRAME{\env}{\chk{\hl}{\AEQ}{t'_2}{t_1}{t_2}}{\fx}}}
       \end{array}}
      {}
\]

\subsubsection{Let}

To test against a let, do it first:
\[
\infer[\rn{RTlet}]
      {\begin{array}{l}
         \ms{\HH}{\PH}{\env}{\fx}{\chk{\hl}{\AEQ}{\LET{x}{t'_1}{t'_2}}{t_1}{t_2}}\stepsto{\AT} {}\\\quad
         \ms{\HH}{\PH}{\env}{\fx}{\LET{x}{t'_1}{\chk{\hl}{\AEQ}{t'_2}{t_1}{t_2}}}
       \end{array}}
      {}
\]

\subsubsection{Letrec}

To test against a let rec, do it first:
\[
\infer[\rn{RTletrec}]
      {\begin{array}{l}
         \ms{\HH}{\PH}{\env}{\fx}{\chk{\hl}{\AEQ}{\LETREC{x_1=v_1,\ldots,x_n=v_n}{t}}{t_1}{t_2}}\stepsto{\AT} {}\\\quad
         \ms{\HH}{\PH}{\env}{\fx}{\LETREC{x_1=v_1,\ldots,x_n=v_n}{\chk{\hl}{\AEQ}{t}{t_1}{t_2}}}
       \end{array}}
      {}
\]

\subsubsection{Conditionals}

To test against an if, do it first:
\[
\infer[\rn{RTif}]
      {\begin{array}{l}
         \ms{\HH}{\PH}{\env}{\fx}{\chk{\hl}{\AEQ}{\IF{x}{t'_1}{t'_2}{t'_3}}{t_1}{t_2}}\stepsto{\AT} {}\\\quad
         \ms{\HH}{\PH}{\env}{\fx}{\IF{x}{t'_1}{\chk{\hl}{\AEQ}{t'_2}{t_1}{t_2}}{\chk{\hl}{\AEQ}{t'_3}{t_1}{t_2}}}
       \end{array}}
      {}
\]

\subsubsection{Stage}

To test against a stage, test against the second subterm:
\[
\infer[\rn{RTlet}]
      {\ms{\HH}{\PH}{\env}{\fx}{\chk{\hl}{\AEQ}{\stage{\fx'}{d}{t'_1}{t'_2}}{t_1}{t_2}}\stepsto{\AT}
       \ms{\HH}{\PH}{\env}{\fx}{\chk{\hl}{\AEQ}{t'_2}{t_1}{t_2}}}
      {}
\]

\omitthis{
\[
\infer[\rn{RT}]
      {\ms{\HH}{\PH}{\env}{\fx}{\chk{\hl}{\AEQ}{}{t_1}{t_2}}\stepsto{\AT}
       \ms{\HH}{\PH}{\env}{\fx}{}}
      {}
\]
}

\subsubsection{Head Labels}

Testing a head label against another head label assumed to be equal to
it, succeeds:
\[
\infer[\rn{RThl}]
      {\ms{\HH}{\PH}{\env}{\fx}{\chk{\hl_1}{\AEQ}{\hl_2}{t_1}{t_2}}\stepsto{\AT}
       \ms{\HH}{\PH}{\env}{\fx}{t_1}}
      {(\hl_1,\hl_2)\in\AEQ}
\]

Testing a head label against another where the first is an integer,
checks that the second is that integer:
\[
\infer[\rn{RThli1}]
      {\ms{\HH}{\PH}{\env}{\fx}{\chk{\hl_1}{\AEQ}{\hl_2}{t_1}{t_2}}\stepsto{\AT}
       \ms{\HH}{\PH}{\env}{\fx}{t_1}}
      {(\hl_1,\hl_2)\notin\AEQ & \HH(\hl_1)=i & \HH(\hl_2)=i}
\]
\[
\infer[\rn{RThli2}]
      {\ms{\HH}{\PH}{\env}{\fx}{\chk{\hl_1}{\AEQ}{\hl_2}{t_1}{t_2}}\stepsto{\AT}
       \ms{\HH}{\PH}{\env}{\fx}{t_2}}
      {(\hl_1,\hl_2)\notin\AEQ & \HH(\hl_1)=i & \hl_2\in\dom\HH & \neg(\HH(\hl_2)=i)}
\]

Testing a head label against another head label where the first is a
table, checks that the second is a table with the same domain, and
then compares the corresponding element head labels consequtively
assuming the outer head labels are equal:
\[
\infer[\rn{RThltab1}]
      {\ms{\HH}{\PH}{\env}{\fx}{\chk{\hl_1}{\AEQ}{\hl_2}{t_1}{t_2}}\stepsto{\AT}
       \ms{\HH}{\PH}{\env}{\fx}{t}}
      {\begin{array}{c}
         (\hl_1,\hl_2)\notin\AEQ \\
         \HH(\hl_1)=\TABLE{\tabeV{i_1}{\hl_{11}},\ldots,\tabeV{i_n}{\hl_{1n}}} \\
         \HH(\hl_1)=\TABLE{\tabeV{i_1}{\hl_{21}},\ldots,\tabeV{i_n}{\hl_{2n}}} \\
         t=\chk{\hl_{11}}{\AEQ\cup\litset{(\hl_1,\hl_2)}}{\hl_{21}}
           {\cdots\chk{\hl_{1n}}{\AEQ\cup\litset{(\hl_1,\hl_2)}}{\hl_{2n}}{t_1}{t_2}\cdots}{t_2} \\
       \end{array}}
\]
\[
\infer[\rn{RThltab2}]
      {\ms{\HH}{\PH}{\env}{\fx}{\chk{\hl_1}{\AEQ}{\hl_2}{t_1}{t_2}}\stepsto{\AT}
       \ms{\HH}{\PH}{\env}{\fx}{t_2}}
      {\begin{array}{c}
         (\hl_1,\hl_2)\notin\AEQ \\
         \HH(\hl_1)=\TABLE{\tabeV{i_1}{\hl_{1}},\ldots,\tabeV{i_n}{\hl_{n}}} \\
         \hl_2\in\dom\HH \\
         \neg(\Histabf{i_1,\ldots,i_n}{\HH(\hl_2)}) \\
       \end{array}}
\]

Testing a head label against another head label where the first is a
closure and the second is not, fails:
\[
\infer[\rn{RThlfun}]
      {\ms{\HH}{\PH}{\env}{\fx}{\chk{\hl_1}{\AEQ}{\hl_2}{t_1}{t_2}}\stepsto{\AT}
       \ms{\HH}{\PH}{\env}{\fx}{t_2}}
      {(\hl_1,\hl_2)\notin\AEQ & \HH(\hl_1)=\clos{\env'}{f} & \hl_2\in\HH & \neg(\Hisfun{\HH(\hl_2)})}
\]

Testing a head label against another head label where the first is a
pointer label, checks that the second is that pointer label:
\[
\infer[\rn{RThlpl1}]
      {\ms{\HH}{\PH}{\env}{\fx}{\chk{\hl_1}{\AEQ}{\hl_2}{t_1}{t_2}}\stepsto{\AT}
       \ms{\HH}{\PH}{\env}{\fx}{t_1}}
      {(\hl_1,\hl_2)\notin\AEQ & \HH(\hl_1)=\pl & \HH(\hl_2)=\pl }
\]
\[
\infer[\rn{RThlpl2}]
      {\ms{\HH}{\PH}{\env}{\fx}{\chk{\hl_1}{\AEQ}{\hl_2}{t_1}{t_2}}\stepsto{\AT}
       \ms{\HH}{\PH}{\env}{\fx}{t_2}}
      {(\hl_1,\hl_2)\notin\AEQ & \HH(\hl_1)=\pl & \hl_2\in\dom\HH & \neg(\HH(\hl_2)=\pl)}
\]

Testing a head label against another where both are closures, commits
an error:
\[
\infer[\rn{RThlfunE}]
      {\ms{\HH}{\PH}{\env}{\fx}{\chk{\hl_1}{\AEQ}{\hl_2}{t_1}{t_2}}\stepstoE}
      {(\hl_1,\hl_2)\notin\AEQ & \HH(\hl_1)=\clos{\env_1}{f_1} & \HH(\hl_2)=\clos{\env_2}{f_2}}
\]

Testing a head label against another head label, where one is
undefined, commits an error:
\[
\infer[\rn{RThlE}]
      {\ms{\HH}{\PH}{\env}{\fx}{\chk{\hl_1}{\AEQ}{\hl_2}{t_1}{t_2}}\stepstoE}
      {(\hl_1,\hl_2)\notin\AEQ & \hl_1\notin\dom\HH\lor\hl_2\notin\dom\HH}
\]

\subsection{Programs}

Programs themselves can terminate with some sequence of actions
performed, can diverge with a (infinite) sequence of actions
performed, or commit a type error.  The judgements and rules for
these are as follows.

A program terminates with a sequence of actions if the initial machine
state makes a finite sequence of steps with those actions to a
terminal machine state denoting an empty table:
\[
\infer[\rn{RP1}]
      {t\evalsto\langle a_1,\ldots,a_n\rangle}
      {\ms{\emptyseq}{\emptyseq}{\emptyseq}{\fxA}{t}\stepsto{a_1}\cdots\stepsto{a_n}
         \ms{\HH}{\PH}{\emptyseq}{\fxA}{\hl} &
       \HH(\hl)=\TABLE{}}
\]

A program diverges with an infinite sequence of actions if the initial
machine state makes an infinite sequence of steps with those actions:
\[
\infer[\rn{RP2}]
      {t\diverges\langle a_1,\ldots\rangle}
      {\ms{\emptyseq}{\emptyseq}{\emptyseq}{\fxA}{t}\stepsto{a_1}\cdots}
\]

A program that reduces to a terminal state that is not an empty table
commits an error:
\[
\infer[\rn{RPE1}]
      {t\evalstoE}
      {\ms{\emptyseq}{\emptyseq}{\emptyseq}{\fxA}{t}\stepsto{a_1}\cdots\stepsto{a_n}
         \ms{\HH}{\PH}{\emptyseq}{\fxA}{\hl} &
       \neg(\HH(\hl)=\TABLE{})}
\]

A program that fails at the toplevel commits an error:
\[
\infer[\rn{RPE2}]
      {t\evalstoE}
      {\ms{\emptyseq}{\emptyseq}{\emptyseq}{\fxA}{t}\stepsto{a_1}\cdots\stepstoF}
\]

A program that commits an error during reduction commits an error:
\[
\infer[\rn{RPE3}]
      {t\evalstoE}
      {\ms{\emptyseq}{\emptyseq}{\emptyseq}{\fxA}{t}\stepsto{a_1}\cdots\stepstoE}
\]

\section{Properties}

The runtime is defined careful to satisfy several important
properties.

First, the head heap only grows monotonically, the pointer heap only
grow monotonically, although the contents of each pointer can change,
and the environment and allowed effects do not change.

\begin{thm}
If $\ms{\HH_1}{\PH_1}{\env_1}{\fx_1}{t_1}\stepsto{a}\ms{\HH_2}{\PH_2}{\env_2}{\fx_2}{t_2}$ then:
\[
\begin{array}{l}
\HH_2=\HH_1,\HH'_2 \\
\forall\pl=\ptrc{\env}{t}{\hl}\in\PH_1:\exists\hl':\PH_2(\pl)=\ptrc{\env}{t}{\hl'} \\
\env_2=\env_1 \\
\fx_2=\fx_1 \\
\end{array}
\]
\end{thm}
\begin{prf}
The proof is by induction on the derivation of the reduction and
inspection of the rules.  All the cases are straightforward.
\end{prf}

Second, the rules are deterministic with respect to the input integers.

\begin{thm}
\label{thm:unique-reduction}
For any $M$ exactly one of the following holds:
\begin{itemize}
\item $M=\ms{\HH}{\PH}{\env}{\fx}{\hl}$ for some $\HH$, $\PH$, $\env$, $\fx$, and $\hl$,
\item $M\stepsto{a}M'$ and if the next input integer is fixed then $a$ and $M'$ are determined by $M$,
\item $M\stepstoF$, or
\item $M\stepstoE$.
\end{itemize}
\end{thm}
\begin{prf}
Let $M$ be $\ms{\HH}{\PH}{\env}{\fx}{t}$ and fix the next input
integer.  The proof is by induction on the structure of $t$ and
inspection of the cases for $t$, the rules, and consideration of cases
of variables in environments, head labels in head heaps, and pointer
labels in pointer heaps.  All the cases are straightforward.
\end{prf}

\begin{thm}
For any source term $t$ if the input integers are fixed then exactly one of the following holds:
\begin{itemize}
\item $t\evalsto\langle a_1,\ldots,a_n\rangle$ for uniquely determined $a_1$, \ldots, $a_n$,
\item $t\diverges\langle a_1,\ldots\rangle$ for uniquely determined $a_1$, \ldots, or
\item $t\evalstoE$.
\end{itemize}
\end{thm}
\begin{prf}
Let $t$ be a source term.  By induction and
Theorem~\ref{thm:unique-reduction}, any two reduction sequences from
$\ms{\emptyseq}{\emptyseq}{\emptyseq}{\fxA}{t}$ must be equal to any
finite point in the reduction sequence.  Therefore, there is a unique
maximal reduction from that state.  If that sequence is infinite then
only rule $\rn{RP2}$ is applicable, and so the second case above holds
and only that case holds.  Otherwise the sequence is finite and rule
$\rn{RP2}$ is not applicable and the second case above does not hold.
By Theorem~\ref{thm:unique-reduction}, either the final machine state
has the form $\ms{\HH}{\PH}{\emptyseq}{\fxA}{\hl}$ for some $\HH$, $\PH$,
and $\hl$; or the sequence ends in failure or error (and only one of
these three cases holds).  For the first, either $\HH(\hl)$ is
$\TABLE{}$ or not, in which case, only rule $\rn{RP1}$ or $\rn{RPE1}$
is applicable and cases one and three above hold and only hold
respectively.  For the other two, only rule $\rn{RPE2}$ or $\rn{RPE3}$
is applicable and case three above holds and only holds.  Also by
Theorem~\ref{thm:unique-reduction}, the actions are uniquely
determined by $t$ and the input integers.
\end{prf}

\bibliographystyle{alpha}
\bibliography{bib}
\end{document}